\documentclass[10pt, twocolumn]{IEEEtran}
\usepackage[lined,commentsnumbered]{algorithm2e}
\usepackage{etoolbox}
\usepackage{cite}
\usepackage{epsfig}
\usepackage{epstopdf}
\usepackage{amsmath,amssymb,amsfonts}
\usepackage{mathrsfs}
\usepackage{framed}
\usepackage{courier}
\usepackage{algpseudocode}
\usepackage{graphicx}
\usepackage{amsthm}
\usepackage[normalem]{ulem}
\usepackage{textcomp}
\usepackage{booktabs}
\usepackage[utf8]{inputenc}
\usepackage{authblk}

\usepackage{mathtools}
\usepackage[dvipsnames]{xcolor}

\begin{document}

%\title{\huge{UAV Connectivity and Energy  Efficiency High performance for  Reliable Communication in a Harsh environment }}

\title{\huge{ Climate-Resilient UAVs: Enhancing Energy-Efficient B5G Communication in Harsh Environments }}
\author{
    Abdu Saif$^{1}$,
    Saeed Hamood Alsamhi$^{2}$,
    Edward Curry$^{2}$\\
    $^{1}$Faculty of Engineering, Taiz University, Taiz, Yemen\\
    \textit{Email:} saif.abduh2016@gmail.com\\
    $^{2}$Insight Centre for Data Analytics, University of Galway, Galway, Ireland\\
    \textit{Emails:} saeed.alsamhi@insight-centre.org, edward.curry@insight-centre.org
}
\maketitle
\vspace{-0.5in}
\begin{abstract}{
The deployment of Beyond Fifth Generation (B5G) networks is increasingly vital yet challenging due to harsh environmental conditions exacerbated by climate change. Unmanned Aerial Vehicles (UAVs) have emerged as critical enablers for B5G communication in adverse weather conditions, including rain, fog, and snow. This paper investigates the synergy between climate-resilient UAVs and energy-efficient B5G communication. We assess UAV coverage and energy efficiency across varying elevation angles in adverse weather. Our findings highlight the pronounced impact of rainfall on UAV coverage and the substantial influence of fog and snow on communication dynamics. Our paper unveils significant improvements in energy efficiency attributed to reduced interference, enhanced data transmission rates, and optimal channel gain under diverse weather conditions. This paper addresses the challenges of harsh environments, emphasizing the potential of climate-resilient UAVs to enhance energy-efficient B5G communication. It underscores the importance of technology in mitigating climate change's impact on communication systems, charting a path toward a more sustainable and resilient future. }
\end{abstract} 

%\newpage

\section{Introduction}

The deployment of Beyond Fifth Generation (B5G) networks represents a pivotal advancement in contemporary communication systems, poised to usher in an era of unparalleled connectivity and capabilities. The networks are designed to transcend the boundaries of their predecessors, offering revolutionary enhancements in data rates, network capacity, and latency reduction \cite{1}. The promises of B5G networks are far-reaching, encompassing applications that range from immersive augmented reality experiences to seamless Internet of Things (IoT) connectivity and mission-critical communications for autonomous vehicles and industrial automation\cite{saif2021internet}.  One of the most pressing issues is the increasing prevalence and severity of adverse environmental conditions, exacerbated by the changing climate. The effects of the conditions, such as heavy rainfall, dense fog, and snowfall, pose formidable obstacles to the reliable operation of wireless communication systems \cite{2}.  Rain, for instance, can attenuate radio signals, leading to signal degradation and reduced coverage area. Similarly, fog can scatter signals, causing signal loss and impairing the quality of communication links. In snowy conditions, the accumulation of ice and snow on communication equipment can disrupt signal propagation and, in some cases, cause equipment failure.
In response to these challenges, Unmanned Aerial Vehicles (UAVs) have emerged as a disruptive technology, offering a promising solution to enhance communication in harsh environments. UAVs can swiftly navigate and adapt to challenging meteorological conditions, thus making UAVs ideal candidates for providing coverage and support to GNs in need \cite{3},\cite{saif2023skyward} offering an up-and-coming solution to enhance communication in harsh environmental conditions. UAVs have demonstrated exceptional adaptability and agility, which make them ideal candidates for providing critical coverage and support to GNs operating in challenging environments \cite{3}. Unlike traditional fixed communication infrastructure, UAVs can autonomously reposition themselves in real-time to optimize signal transmission and reception, ensuring uninterrupted communication even in adverse weather phenomena such as heavy rain, thick fog, or snowstorms. This capability significantly enhances the reliability of communication networks in harsh environments, where maintaining a consistent and high-quality connection is paramount \cite{3}. UAVs can significantly reduce routing overhead. Traditional GNs in harsh environments often rely on complex routing schemes to circumvent obstacles and maintain connectivity. With UAV's direct line-of-sight capability and adaptability, UAVs streamline routing and reduce the complexity of data transmission, leading to more efficient network operation \cite{4}.

Moreover, UAV-assisted cellular communication represents a vital technology for fulfilling the dynamic communication requirements of modern society\cite{saif2021unmanned}. As traditional communication infrastructure is susceptible to malfunction and disruption during natural disasters and extreme weather events \cite{55},\cite{al2022d2d} UAVs provide swift recovery, ensuring that communication services remain available despite adversity\cite{abo2022adaptive}. Furthermore, the adaptability and resilience of UAVs extend to situations where communication infrastructure is temporarily overloaded or incapacitated due to unexpected surges in network traffic, as often occurs during major events or emergencies. In these cases, UAVs can be rapidly deployed to offload traffic, ensuring that critical communication services, including emergency calls and data transmission, continue to operate smoothly \cite{66}.
This paper delves into the intersection of UAVs, adverse meteorological conditions, and B5G communication, evaluating the role of UAVs in enhancing energy-efficient communication amidst rain, fog, and snow. The assessment encompasses a range of performance metrics, including energy efficiency, outage probability, spectrum efficiency, and path loss. By addressing these critical aspects, we aim to shed light on the potential of UAV-assisted systems to thrive in harsh environments and to serve as viable replacements for dysfunctional GNs in such scenarios.

\subsection{Motivation and contributions}

The advent of B5G networks promises revolutionary advancements in communication, enabling applications like ultra-low latency for autonomous vehicles and seamless connectivity for IoT devices. The impact of harsh environmental conditions brought on by climate change, such as heavy rain, fog, and snow, which present significant obstacles to dependable wireless communication, is limiting the effectiveness of B5G networks. This paper aims to bridge this gap by exploring the role of climate-resilient UAVs in enhancing energy-efficient B5G communication in adverse weather conditions, unlocking the full potential of B5G networks for a wide range of transformative applications.
While B5G networks hold immense promise, a critical gap exists in ensuring the resilience of communication infrastructure to the adverse meteorological conditions intensified by climate change. Rain, fog, and snow can severely degrade wireless communication quality, disrupting services and hindering B5G network deployment. This paper fills this gap by looking into how climate-resilient UAVs can be used to improve energy-efficient B5G communication in harsh environments. It offers crucial insights and solutions for keeping communication services running in bad weather, ensuring critical applications do not stop working, and making B5G communication systems more climate-resilient.

The paper aims to advance understanding of how UAVs can play a pivotal role in enhancing the resilience and energy efficiency of B5G communication networks in harsh environmental conditions. The contributions encompass demonstrating UAVs as a climate-resilient solution and quantifying their impact on energy efficiency and key performance metrics. The contributions of the paper are summarised as follows:

\begin{itemize}
\item We introduce climate-resilient UAVs as a significant contribution to the field. It showcases how UAVs can be dependable for maintaining communication services in adverse meteorological conditions (i.e., rain, fog, and snow). The paper highlights UAVs as a critical element in addressing the challenges posed by climate-induced weather conditions in communication infrastructure.
\item The paper quantifies and emphasizes the significant energy efficiency improvements made possible by integrating UAVs into B5G communication networks in harsh environments. The contributions highlight the reduced interference, augmented data transmission rates, and optimal channel gain facilitated by UAVs, underscoring their potential to optimize energy consumption in communication systems.

\item We provide a holistic assessment of the proposed UAV assisted system by evaluating key performance metrics. By addressing energy efficiency, outage probability, spectrum efficiency, and path loss, it offers a comprehensive understanding of the capabilities and limitations of UAVs in mitigating the adverse effects of weather on B5G communication. 
\end{itemize}
\subsection{Related work}
The intersection of UAVs, adverse meteorological conditions, and wireless communication has garnered considerable attention in recent research. Several studies have laid the foundation for understanding the potential of UAVs in mitigating the challenges posed by harsh environments.
The authors of \cite{4} conducted pioneering research on wireless communications with UAVs, elucidating the opportunities and challenges of UAV-assisted communication. The authors highlighted the agility and adaptability of UAVs in providing on-demand wireless links, particularly in scenarios where traditional infrastructure is compromised\cite{saif2021infrastructure}. While their study focused on general UAV applications, it provided valuable insights into the feasibility of UAVs in adverse environmental conditions. In \cite{3},  presented a comprehensive tutorial on the applications, challenges, and open problems associated with UAVs in wireless networks. The authors of \cite{r3} delved into the effects of fog and haze on visible light communication. While not UAV-centric, their work highlighted the need for resilient communication solutions in adverse weather. The study emphasized the impact of weather conditions on communication quality and reliability, aligning with our research objectives.

The authors of \cite{r44} explored the utilization of UAVs in 6G communication networks, emphasizing their ability to provide rapid deployment and coverage extension during extreme weather events. In \cite{r4} investigated the impact of rainfall on millimetre-wave communication, a key component of B5G and 6G networks. The findings highlighted the significance of weather-induced signal attenuation and the potential for UAVs to act as dynamic relays, mitigating the effects of rain on high-frequency communication. Furthermore, \cite{r5} delved into the challenges of energy-efficient communication in adverse weather. The work specifically examined energy-aware routing protocols for UAV-assisted networks operating in foggy conditions, shedding light on the importance of energy efficiency in challenging meteorological contexts.

While the above studies have made significant strides in understanding the role of UAVs and weather effects in communication networks, our paper contributes to this evolving landscape by conducting a comprehensive assessment of UAV performance in B5G networks under the combined influence of rain, fog, and snow. Our paper provides insights into energy efficiency, outage probability, spectrum efficiency, and path loss, addressing specific challenges related to climate-resilient and energy-efficient B5G communication in harsh environments.

%\subsection{Paper Structure}

\section{Proposed System Model}

The system model is tailored to address a scenario where GNs face formidable challenges in maintaining wireless connectivity, primarily due to adverse environmental conditions, encompassing rain, fog, snow, and other disruptive factors. These conditions disrupt the conventional wireless coverage services typically provided by Ground Base stations, necessitating an innovative approach. We strategically deploy UAVs to reinstate and sustain essential communication links with GNs in these demanding scenarios to overcome this pervasive issue. 
In our proposed model, UAVs have advanced directional antennas, a critical feature designed to optimize network coverage performance. Furthermore, the UAVs employ dynamic altitude adjustments, responding to factors such as antenna beamwidth and the density of nearby structures, as quantified by the number of installations \cite{saif2023flexible}. 

The altitude adaptation ensures efficient GN coverage, especially in harsh environmental conditions. The UAVs' capability to dynamically allocate GNs and user devices within the coverage area significantly ensures reliable and uninterrupted connectivity in challenging scenarios. The allocation strategy enhances coverage and connectivity, even in adverse weather conditions.
To assess the efficacy of our proposed system, we employ a range of performance parameters, including path loss, energy efficiency, and coverage area. These metrics are vital for evaluating the system's performance across diverse weather conditions, ensuring reliable connectivity is maintained in the harshest environments.

\subsection{Attenuation Models for Rain, Fog and Snow }
 Utilizing UAVs for telecommunications services in adverse weather conditions presents a significant challenge. To establish a foundation for deploying UAV communications in such dynamic environments, it is essential to carefully investigate attenuation models for various typical weather conditions, including rain, fog, and snow. The attenuation models for these weather conditions are expressed as follows:

 \begin{align}
  {\gamma } = \begin{dcases*}
k{R^\alpha }& Rain model\\
{K_1}(f,T)M\quad ({\text{dB}}/{\text{km}})& Fog model\\
        0.00349\frac{{R_{\text{s}}^{1.6}}}{{{\lambda ^4}}} + 0.00224\frac{{{R_{\text{s}}}}}{\lambda }\quad ({\text{dB}}/{\text{km}})&Snow model 
       \end{dcases*}
       \label{Eq6}
  \end{align}

\textbf{Rain Model:} The rain attenuation model is characterized by the equation ${\gamma} = k{R^\alpha}$, where $R$ represents the rain rate in millimetres per hour exceeded for 0.01% of an average year. The parameters $k$ and $\alpha$ are functions of polarization, and their values are determined by equations derived from experimental data fitting to power-rate coefficients accessible from ITU-R~\cite{mello2007prediction}.

\textbf{Snow Model:} The snow attenuation model is described by the equation $0.00349\frac{{R_{\text{s}}^{1.6}}}{{{\lambda ^4}}} + 0.00224\frac{{{R_{\text{s}}}}}{\lambda }$, where $\lambda$ is the wavelength and $R_{\text{s}}$ is the snowfall speed.

\textbf{Fog Model:} The attenuation coefficient is given by $\frac{{4.34 f^2}}{{\lambda^2}}$ in units of $(dB/km)/(g/m^3)$. Where,

$\eta$ is defined as $\frac{2+\varepsilon ^{\prime } }{\varepsilon ^{\prime \prime } }$, and the complex permittivity of water is expressed as ${\varepsilon ^{\prime \prime }}(f)$ and ${\varepsilon ^\prime }(f)$.
For fog attenuation modelling, the equations governing the complex permittivity of water (${\varepsilon ^{\prime \prime }}(f)$ and ${\varepsilon ^\prime }(f)$) are provided, with temperature ($T$) and various constants determining their values. The specific attenuation coefficient (${K_1}(f, T)$) is essential for quantifying fog-induced attenuation and is directly related to the density of liquid water in the cloud or fog ($M$).

\begin{equation*}{K_1}(f,T) = \frac{{0.819f}}{{{\varepsilon ^{\prime \prime }}\left( {1 + {\eta ^2}} \right)}}\quad ({\text{dB}}/{\text{km}})/\left( {{\text{g}}/{{\text{m}}^3}} \right),\tag{2}\end{equation*}

where $\eta=\frac{2+\varepsilon ^{\prime } }{\varepsilon ^{\prime \prime } }$ and the $\varepsilon ^{\prime \prime }(f)$ is given as: 

\begin{align*} & {\varepsilon ^{\prime \prime }}(f) = \frac{{f\left( {{\varepsilon _0} - {\varepsilon _1}} \right)}}{{{f_p}\left[ {1 + {{\left( {f/{f_p}} \right)}^2}} \right]}} + \frac{{f\left( {{\varepsilon _1} - {\varepsilon _2}} \right)}}{{{f_s}\left[ {1 + {{\left( {f/{f_s}} \right)}^2}} \right]}},\tag{3} \\ & {\varepsilon ^\prime }(f) = \frac{{{\varepsilon _0} - {\varepsilon _1}}}{{\left[ {1 + {{\left( {f/{f_p}} \right)}^2}} \right]}} + \frac{{{\varepsilon _1} - {\varepsilon _2}}}{{\left[ {1 + {{\left( {f/{f_s}} \right)}^2}} \right]}} + {\varepsilon _2},\tag{4}\end{align*}

Where $\varepsilon_0$ is calculated as $77.66 + 103.3(\theta - 1)$, $\varepsilon_1$ is derived as $0.0671$ times $\varepsilon_0$, $\varepsilon_2$ is a constant with a value of $3.52$. Here, $\theta$ is determined as $300/T_{\text{fog}}$, where $T_{\text{fog}}$ represents the temperature during foggy weather conditions and is set to 293.15K. Additionally, we define the primary relaxation frequency, $f_p$, as $20.20 - 146(\theta - 1) + 316(\theta - 1)^2$, and the secondary relaxation frequency, $f_s$, as $39.8$ times $f_p$ (in GHz).

%where $m$ and $\alpha$ , are determined  \cite{song2020meteorologically},
$T$ represents the temperature of liquid water, $K_{1}$ stands for the specific attenuation coefficient ($\text{dB/km}$ per $\text{g/m}^3$), and $M$ denotes the density of liquid water in the cloud or fog ($\text{g/m}^3$), as documented in \cite{zang2019impact}. Additionally, $Rs$ corresponds to the snowfall speed measured in millimeters per hour, and $\lambda$ indicates the wavelength measured in centimeters

 The attenuation models are crucial for understanding and mitigating the effects of rain, fog, and snow on wireless communication, particularly when deploying UAVs in challenging meteorological environments, providing valuable insights into how environmental conditions impact communication performance and guide the development of resilient communication systems.

\subsection{Path Loss for Rain, Fog and Snow}

Path loss propagation plays a pivotal role in shaping the wireless communication channel between UAVs and GNs within the air-to-ground (A2G) channel. Environmental parameters, including the distance between UAVs and GNs, GN elevation angles, and UAV altitudes, profoundly influence the signal path loss. To comprehensively model the A2G channel under multiple weather conditions, we integrate the specific attenuation coefficients for rain, fog, and snow into the path loss models. The resulting A2G channel models can be expressed as:\\

\begin{equation*}\begin{array}{c} P{L_{{\text{UAV}}}} = \left( {P{L_{{\text{LoS}}}} \times {P_{{\text{LoS}}}} + P{L_{{\text{NLoS}}}} \times {P_{{\text{NLoS}}}}} \right) + \frac{{(\beta + \gamma )d}}{{1000}} \\ = \left( {\frac{A}{{1 + a\exp \left( { - b\left( {\frac{{180}}{\pi }{{\tan }^{ - 1}}\left( {\frac{h}{r}} \right) - a} \right)} \right)}}} \right. \\ \left. { + 20\log \frac{r}{{\cos \left( {\frac{{180}}{\pi }{{\tan }^{ - 1}}\left( {\frac{h}{r}} \right)} \right)}} + B} \right) + \frac{{(\beta + \gamma )d}}{{1000}},\end{array} \tag{5}\end{equation*}

Where, $PL_{\text{UAV}}$ represents the UAV path loss, $PL_{\text{LoS}}$ and $PL_{\text{NLoS}}$ are path loss components for Line-of-Sight (LoS) and Non-Line-of-Sight (NLoS) conditions, respectively. The variables $A$ and $\gamma$ account for specific attenuation coefficients related to various weather scenarios in dB/km. Additionally, $f_c$ denotes the carrier frequency, $c$ is the speed of light, and $d_{[ki]}$ represents the distance from the UAV to the GNs. The terms $\eta_{\text{LoS}}$ and $\eta_{\text{NLoS}}$ consider excessive loss due to shadowing and scattering in LoS and NLoS links, while $\beta$ represents atmospheric attenuation.

\subsection{Connectivity in UAV coverage area}

To determine the optimal coverage area of the UAV, we derive the first derivative of equation (16) concerning the variable $r$ as follows:

\begin{equation*}
\begin{aligned}
0 &= -\frac{Aah}{r^2} \left( b \right) \left( \tan^{-1} \left( \frac{h}{r} \right) - a \right) \\
&\quad \times \left( e^{-b \left( \tan^{-1} \left( \frac{h}{r} \right) - a \right) } \right) \\
&\quad \times \left( 1 + a e^{-b \left( \tan^{-1} \left( \frac{h}{r} \right) - a \right) } \right)^{-2} \\
&\quad \times \left( \frac{h^2}{r^2} + 1 \right)^{-1} \\
&\quad + 20 \log \left( \cos \left( 180 \frac{1}{\pi} \arctan \left( \frac{h}{r} \right) \right) \right)^{-1} \\
&\quad - 3600 \frac{\left( \log h \right) h}{r \pi} \sin \left( 180 \frac{1}{\pi} \tan^{-1} \left( \frac{h}{r} \right) \right) \\
&\quad \times \left( \cos \left( 180 \frac{1}{\pi} \arctan \left( \frac{h}{r} \right) \right) \right)^{-2} \\
&\quad \times \left( \frac{h^2}{r^2} + 1 \right)^{-1}
\end{aligned}
\tag{6}
\end{equation*}

This derivative is instrumental in determining the optimal coverage area of the UAV, taking into account critical parameters like altitude, distance, and environmental factors. By findings from \cite{al2014optimal}, we combine the propagation attenuation effects of rain, fog, and snow under various weather conditions to model wireless channels effectively. This integrated model helps us understand how adverse weather impacts communication performance, particularly when deploying UAVs.

The desired performance metrics for assessing UAV-assisted GN communications encompass several crucial aspects, including achieving higher data rates, enhancing energy efficiency, increasing network capacity, and ensuring service availability during harsh environmental events \cite{zhang2018new}. Notably, network capacity is evaluated based on the traffic that can be handled with a minimum bit error rate, especially in challenging conditions.
The derivative-based optimization approach, coupled with a comprehensive understanding of weather-induced attenuation, forms the basis for effective wireless communication system design, focusing on achieving optimal performance metrics in UAV-assisted GN communications, even under adverse environmental conditions.

\subsection{Energy Efficiency}
In harsh environmental communication scenarios, the UAV takes on the crucial role of a temporary communication relay, ensuring the timely and reliable exchange of vital information. However, adverse weather conditions such as rain, fog, and snow significantly impact the UAV's energy resources, particularly during refilling operations. The conditions consume valuable time and deplete the UAV's energy reserves. As a result, an urgent need exists to enhance the energy efficiency of UAV-assisted communication systems.
The holistic evaluation of energy efficiency in the context of UAV-assisted communication involves considering the entire instantaneous transmission vector of the UAV, denoted as $EE^{k}_{UAV}$. This vector encompasses all the constituent elements, representing every link between the UAV and the ground nodes. It serves as a comprehensive metric to assess and optimize the system's energy efficiency, addressing the challenges posed by harsh environmental conditions and ensuring the effective utilization of the UAV's resources for reliable communication.

\begin{equation}
\mbox {EE}^{[k]}_{\text{UAV}}=\frac{B \cdot \mbox{log}_2(1+\frac {p_{i} {h}_{i}}{\sum_{m=1}^{M}p_{m}h_{m,i} + p_{j} {h}_{j} + \sigma^{2}})}{P_{tx}h}\tag{7},
\label{7}
\end{equation}
Where, $h$ represents the number of hops from the UAV to GN communications, and $P_{tx}$ stands for the maximum transmission power used by the UAV for downlink communications with GNs.

\section{Results and Discussion}
In this section, we present an extensive array of simulation results that vividly illustrate the performance of the proposed schemes. Our evaluation encompasses critical parameters and metrics, focusing on LoS probability, path loss, and energy efficiency in UAV-to-ground node communication. These evaluations are conducted across harsh environmental scenarios, including rain, fog, and snow.

Our simulations consider a scenario where ground nodes are randomly distributed within the UAV's coverage area, mirroring real-world conditions. To achieve a comprehensive assessment, we vary ground node elevation angles and adjust UAV altitudes accordingly. Considering the UAV's coverage area, we assume the ground nodes are scattered randomly in each scenario. The transmission distance between the source and destination of the ground nodes and the UAV is set from 100 meters to 1000 meters. Moreover, we explore a wide range of elevation angles for the ground nodes, spanning from 0 degrees to 90 degrees. By conducting rigorous simulations, we aim to provide a comprehensive understanding of how our proposed schemes perform under varying conditions and scenarios. This empirical evidence will validate the efficacy of our approaches and offer insights into their real-world applicability.

\begin{table}[!t]
\caption{Simulation Parameters}
\centering
\renewcommand{\arraystretch}{1.3}
\begin{tabular}{|p{4cm}|p{4cm}|} \hline
\textbf{Parameters} & \textbf{Values} \\ \hline \hline	
B & $5$~MHz \\ \hline                            
$\sigma^2$ \& $ -174$~dBm/Hz \\	 \hline
$f_{c} \& [28~GHz, 60~GHz]$ \\ \hline
Urban area factors & $a = 9.61$, $b=0.16$, $\eta_{\text{LoS}} = 1$, $\eta_{\text{NLoS}} = 20$ \\ \hline
Medium rain & $R = 12.5$~mm/h \\ \hline
Medium fog & $M = 0.05$~g/m$^3$ \\ \hline
Snow & $Rs = 5$~mm/h \\ \hline
Weather conditions [dB/km] attenuation coefficient \cite{mansour2017new} & Clear air = 0.43, Haze = 4.2, \\
& Moderate rain (12.5 mm/h) = 5.8, \\
& Heavy rain (25 mm/h) = 9.2, \\
& Light fog = 20, \\
& Moderate fog = 42.2, \\
& Heavy fog = 125 \\ \hline
UAV altitude & $120$~m \\ \hline
UAV TX power & 5~W \\ \hline  
GN & $0^\circ$ to $90^\circ$ \\ \hline
\end{tabular}
\end{table}

%%%%
%%%%%%%%
\subsection{Path Loss}

\begin{figure}[t!]
	\includegraphics[width=1.1\linewidth]{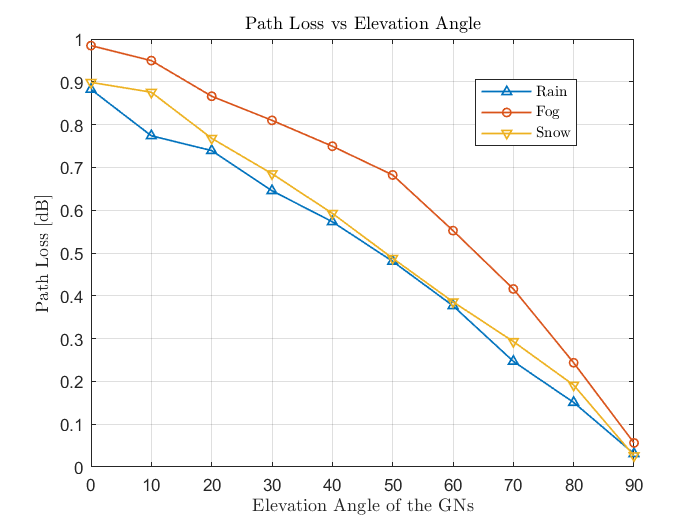}
	\caption{ Variation in Path Loss with GN Elevation Angle in Different Harsh Environments (Rain, Fog, Snow) }
	\label{f4}
\end{figure}

Path loss propagation is a critical consideration when assessing communication performance, and it is essential to understand how it varies under different conditions. In the context of harsh environments, we observed a notable trend in path loss as the elevation angle of the ground nodes (GNs) changed from $0^{\circ}$ to $90^{\circ}$, as depicted in Fig.~\ref{f4}. The variation is primarily attributed to weather conditions such as rain, fog, and snow.
In particular, when GNs are situated in harsh weather conditions, the path loss substantially increases, ranging from 0 dB to 310 dB as the GN elevation angle varies. The dramatic change in path loss indicates the challenges imposed by adverse weather. Notably, in a rainy environment, the path loss experiences a more modest increase, rising from 0 dB to 51 dB. In contrast, the fog environment displays a path loss that escalates from 100 dB to 175 dB. Finally, for the snow environment, path loss undergoes a considerable increase, ranging from 0 dB to 310 dB, as the GN elevation angle varies from $0^{\circ}$ to $90^{\circ}$ due to the specific characteristics of a single city model.

\begin{algorithm}
\caption{Climate-Resilient UAV Communication Model}
\SetAlgoLined
\KwData{
    Rain rate: $rain\_rate\_mm\_hr$ \\
    Frequency: $frequency\_GHz$ \\
    Temperature: $temperature\_Kelvin$ \\
    Snowfall speed: $snowfall\_speed\_mm\_hr$ \\
    Wavelength: $wavelength\_cm$ \\
    $t_{max}$: Maximum number of iterations \\
    $P_{max}$: Maximum transmission power of UAV \\
    $d_{u,i}$: The distance between UAV and GNS \\
    $\theta_{j,k}$: Elevation angles of GNs \\
    $N$: length(weather\_conditions) ('Rain', 'Fog', 'Snow')
}

\KwResult{Energy efficiency and UAV Coverage in conditions of ('Rain', 'Fog', 'Snow')}

\For{$t=1$ to $t_{max}$}{
    \For{$i=1$ to $N$}{
        \For{$j=1$ to $N$}{
            Calculate Path Loss Based on Eq(5) \\
            Calculate UAV Coverage Area Based on Eq(6) \\
            Calculate Energy Efficiency Based on Eq(7)
        }
    }
}
\end{algorithm}

\begin{figure}
	\includegraphics[width=1\linewidth]{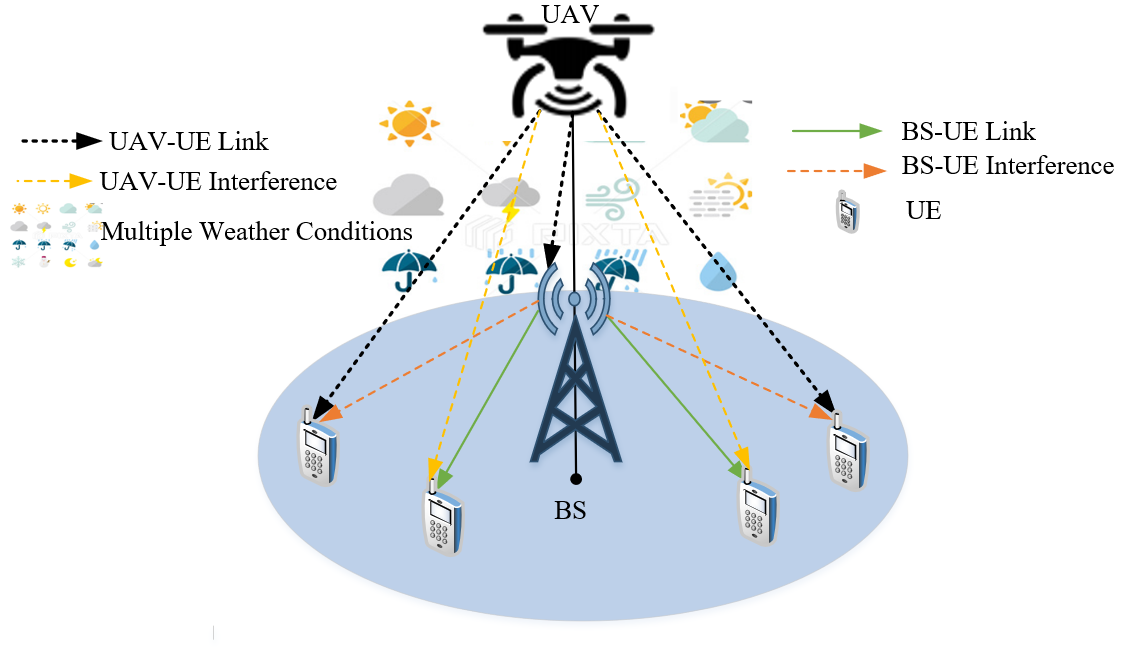}
	\caption{Architecture of the UAV to GN communication in harsh environments}
	\label{f1}
\end{figure}

\subsection{Energy Efficiency}

\begin{figure}[t!]
	\includegraphics[width=1.1\linewidth]{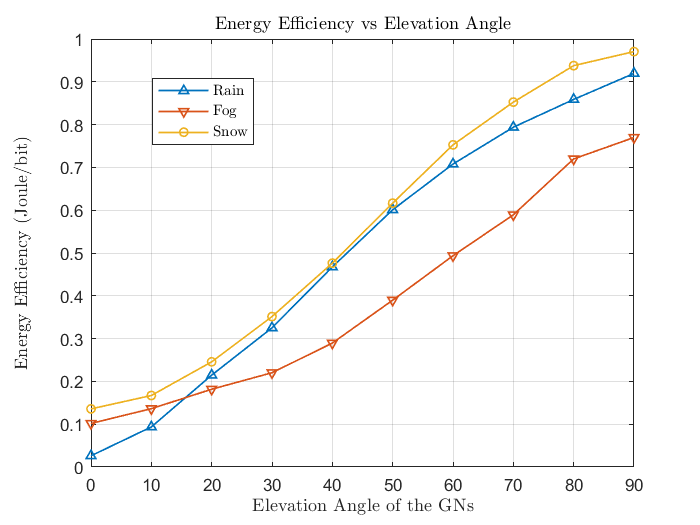}
	\caption {Variation in Energy Efficiency with Ground Node Distance Across Different Scenarios (Rain, Fog, Snow) }
	\label{f5}
\end{figure}

In this context, we delve into the analysis of both the UAV and the GN energy efficiency, shedding light on how these efficiency metrics behave under varying conditions. Fig.~\ref{f5} visually represents our findings.
We observe that energy efficiency experiences a decline as the distance between the GNs and the UAV increases; a decrease in energy efficiency is a significant aspect to consider, as it signifies that more energy is required to maintain communication over longer distances.
However, an exciting observation arises as we examine energy efficiency across different scenarios. As the transmission distance of the ground nodes increases, the energy efficiency values for each scenario start to converge and become closely aligned. The phenomenon suggests that as the ground node distance expands, the impact of interference on energy efficiency diminishes. When GNs are farther apart, interference plays a less significant role in reducing energy efficiency.

Increasing the transmission power is one effective way to counteract this reduction in energy efficiency due to interference, particularly in scenarios with longer transmission distances\cite{saif2023efficient}. The signal can better overcome interference by providing more power, leading to improved energy efficiency.
Furthermore, it is essential to highlight that co-channel interference primarily affects energy efficiency, where multiple communication channels overlap and interfere. In this context, directional antennas can be a valuable strategy. Directional antennas can focus and concentrate the signal in a specific direction, reducing interference and enhancing the overall efficiency metrics. Hence, incorporating directional antennas is a practical approach to mitigate the adverse effects of interference on energy efficiency in UAV-to-GN communication scenarios.

\subsection{UAV Communication Coverage}

\begin{figure}[t!]
	\includegraphics[width=1.1\linewidth]{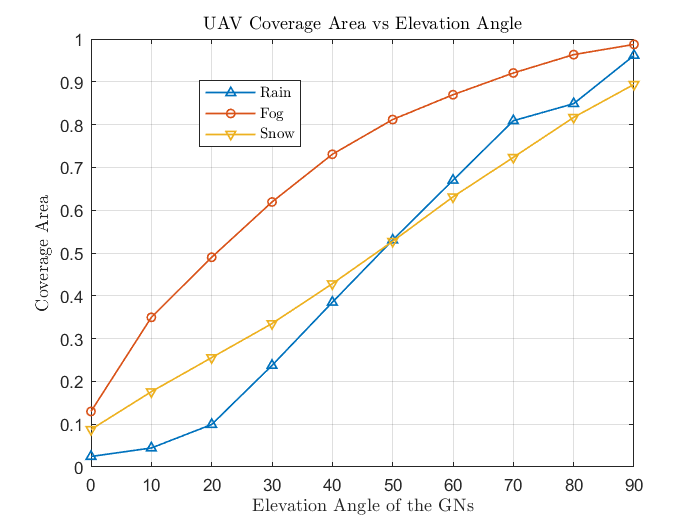}
	\caption{Variation in UAV Communication Coverage Radius with Altitude Across Different Harsh Weather Conditions (Rain, Fog, Snow)}
	\label{f6}
\end{figure}

Fig ~\ref{f6} illustrates the impact of harsh environmental conditions, including rain, fog, and snow, on UAV communication systems' cell radius and operating altitude.  In scenarios with moderate rainfall, the coverage radius and the optimal UAV altitude are relatively small, indicating the need for lower operating altitudes for adequate coverage. Conversely, the coverage radius significantly expands under light snow conditions, enabling UAVs to operate at higher altitudes while providing extensive coverage. This demonstrates that the severity of weather conditions plays a pivotal role in determining the trade-off between coverage area and UAV altitude in UAV communication, with milder conditions allowing for greater altitude and more extensive coverage areas compared to scenarios with more moderate rain.

\section{Conclusion}
In this paper, we have explored the pivotal role of UAV-facilitated ground node communication in challenging meteorological conditions, encompassing rain, fog, and snow. Our assessment has focused on key performance metrics, including energy efficiency, outage probability, spectrum efficiency, and path loss.  In our proposed system model, UAVs offer invaluable coverage support to empower GNs operating in harsh environments, leading to enhanced network scalability, reduced routing overhead, optimized throughput, and expanded coverage. UAV-assisted cellular communication is poised to become a cornerstone technology, catering to the evolving demands of dynamic and diverse communication scenarios. Furthermore, UAVs are indispensable for swift recovery when traditional communication infrastructure succumbs to dysfunction during natural disasters and adverse environmental conditions. Our results affirm the viability of the UAV-assisted system, showcasing its capability to perform on par with GNs in harsh environments. UAVs emerge as a fitting replacement for dysfunctional GNs in these challenging scenarios, offering reliability and rapid restoration of communication services during critical situations.

\section*{Acknowledgment}

This publication has emanated from research conducted with the financial support of Science Foundation Ireland under Grant number SFI/12/RC/2289\_P2.

\bibliography{Referencesbib.bib}
\bibliographystyle{IEEEtran}
\end{document}